\documentclass[8.5pt,twoside,twocolumn]{article}
\usepackage[super,sort&compress,comma]{natbib} 
\usepackage{color}
\usepackage{times,mathptmx}
\usepackage{sectsty}
\usepackage{balance} 
\usepackage{amssymb,amsmath,amsfonts,bm,array}
\usepackage{graphicx} %eps figures can be used instead
\usepackage{lastpage}
\usepackage[format=plain,justification=raggedright,singlelinecheck=false,font=small,labelfont=bf,labelsep=space]{caption} 
\usepackage{hyperref}
\def\p{\partial}
\def\bx{\bgroup \bf x\egroup}
\def\bn{\textbf{n}}
\def\bi{\textbf{e}}
\def\vp{\varphi}
\def\ve{\varepsilon}

\def\const{\mathop {\rm const}\nolimits}

\def\be{\begin{equation}}
\def\ee{\end{equation}}
\let\Im\I
\def\bnu{\bgroup \pmb \nu\egroup}
\def\btau{\bgroup \pmb \tau\egroup}

\newcommand{\mcm}[1]{{\color{black} #1}}
\newcommand{\mcmred}[1]{{\color{black} #1}}

\title{Viral nematics in confined geometries}

\author{O. V. Manyuhina, K. B. Lawlor, M. C. Marchetti and  M. J. Bowick \\[1ex] Physics Department, Syracuse University, Syracuse, NY 13244, USA}

\date{\today}

\begin{document}
\maketitle

\begin{abstract}\bf\small
Motivated by recent experiments on the rod-like virus bacteriophage fd, confined to circular and annular domains, we present a theoretical study of structural  transitions in these geometries. Using the continuum theory of nematic liquid crystals, we examine the competition between bulk elasticity and surface anchoring, mediated by the formation of topological defects. We show analytically that bulk defects are unstable with respect to defects sitting at the boundary. In the case of an annulus, whose topology does not require the presence of topological defects, we find that nematic textures with boundary defects are stable compared to defect-free configurations when the anchoring is weak. Our simple approach, with no fitting parameters, suggests a possible symmetry breaking mechanism responsible for the formation of one-, two- and three-fold textures under annular confinement. 
\end{abstract}

\section*{Introduction}

%experiment
Over the last thirty years rod-like viruses, optically visualizable,  have been established as unique prototype systems for experimental studies of liquid crystalline order~\cite{dogic:review}. In recent experiments, $fd$-viruses $0.88~\mu$m long and $6.6$~nm thick were confined to wedge~\cite{aarts:2012}, rectangular~\cite{aarts:2014} and annular~\cite{thesis:2013} geometries. Depending on the relative size and shape of the confining geometry various two-dimensional (2D) equilibrium configurations of $fd$-viruses with spatially non-trivial orientational texture were observed. The author~\cite{thesis:2013}  attributes the formation of rather striking annular textures with three-fold rotational symmetry  to the finite length of $fd$-rods. Most theoretical work so far has employed numerical Monte Carlo simulations~\cite{aarts:2012,aarts:2014,velasco:2014}, incorporating the range of confinement, the aspect ratio and the density of the constituent rods.   This approach inherently depends on the details of the microscopic parameters. Although very useful in modeling specific small systems these simulations do not provide a universal or complete picture of the phase diagram and corresponding transitions exhibited by confined systems. The aim of this paper is to gain a basic understanding of the symmetry selection mechanisms of nematic configurations confined to circular geometries.

%defects
As suggested by experimental data~\cite{thesis:2013}, the delicate interplay between  boundary and bulk effects is responsible for the variety of equilibrium structures. Confined to thin annuli, $fd$-viruses readily satisfy the preferred tangential alignment at the inner and outer boundaries (planar anchoring) at the expense of bend elastic distortions in the bulk. For a disc, such a bend configuration would inevitably lead to the presence of a topological defect of charge $q=+1$ at the center, in agreement with the Poincar\'e--Hopf theorem~\cite{book:docarmo} applied to line fields, since the Euler characteristic $\chi$ of a disc is one. However, as analytically calculated below, the bulk defect is unstable with respect to two defects sitting at the boundary. As a result  the nematic director satisfies the boundary conditions almost everywhere on the boundary (strong anchoring) except at a finite number of points, similar to the studies of  Langmuir monolayers~\cite{carlson:1988,lubensky:1999}. In case of a finite/weak anchoring, the preferred alignment at the boundary can be violated over extended region. Here we show that even for an annulus boundary defects become energetically favored, though no defects are required by topology ($\chi=0$). By inserting pairs of positive (negative) charges at the outer (inner) boundary of an annulus  we screen the curvature of circumference and  `unbend' otherwise bend nematic texture, similar to the stabilizing effect of the Gaussian curvature in the two-dimensional manifold~\cite{bowick:review,giomi:2008}. Alternatively, introducing boundary defects into the uniform state (no bend or splay elastic deformations), we can curve the director field to satisfy the boundary conditions and lower the anchoring energy.  Therefore, whenever bulk and boundary energy contributions become equally important, we expect the equilibrium  2D textures to encompass boundary defects.

Interestingly, the phenomenon of the transformation between surface and bulk defects was studied in 3D nematic droplets~\cite{volovik:1983}. When the  boundary conditions, set by the temperature, change from the homeotropic to planar anchoring, the nematic director deforms from  the `hedgehog' configuration (point defect inside) to the pair of `boojums', sitting at the surface. In experiments with the $fd$-virus~\cite{thesis:2013}, the nature of the boundaries does not change, while it is the size and the topology (disc or annulus) of the 2D confinement which may influence the distribution of the topological charge at the boundary. To quantify the effect of the confinement and weak anchoring  conditions, we adopt the continuum theory of nematic liquid crystals~\cite{book:intro},  assuming the one elastic constant approximation, such that bend and splay elastic constants are equal $K$  as suggested by~\cite{aarts:2012} for $fd$-virus. We quantify the relative stability and the equilibrium number and charge of defects at the boundary, which is governed by the ratio of the anchoring extrapolation length $L_a=K/W_a$ ($W_a$ is the anchoring strength)  and the system size $R$. With no fitting parameters we capture some features of experiments~\cite{thesis:2013} on $fd$-viruses, namely the appearance of one- and two-fold textures, in particular. Our analytic calculations for different geometries suggest the value of the anchoring extrapolation length $L_a\simeq 5~\mu$m.

In the following we first formulate the problem and characterize topological defects in the bulk and at the boundary. Next we consider special solutions for nematic configurations, confined to the disc and to the annulus, allowing for the presence of defects at the boundary. This approach is merely an  {\it ansatz},  the limiting case of a general minimization problem.

%%%%%%%%%%%%%%%%%%%%%%%%%%%%%%%%%%%%
\section*{Formulation of the problem}
%%%%%%%%%%%%%%%%%%%%%%%%%%%%%%%%%%%%

\begin{figure}[tb]
\centering
\includegraphics[width=\linewidth]{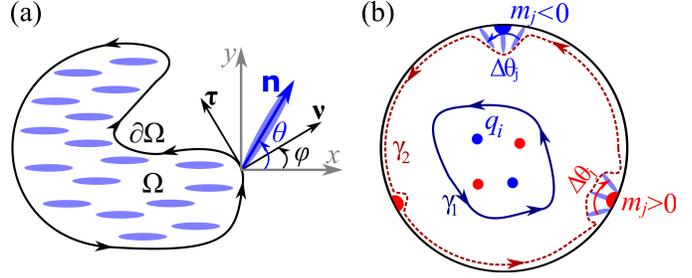}
\caption{\label{fig:charge} Schematic illustration of (a) a uniform line field $\bn=\bi_x$ confined to a simple region $\Omega$ with $\chi=1$~\eqref{eq:chi}, where we have chosen as positive the counter-clockwise orientation of the boundary $\p\Omega$, with the unit normal $\bnu$ pointing outside $\Omega$; (b) two kinds of topological defects:  bulk defects (circles of charge $q_i$~\eqref{eq:qi}) and  boundary defects (half-circles of charge $m_j=\Delta\theta_j/(2\pi)$),  related via the charge conservation law, Eq.~\eqref{eq:conserv}.}
\end{figure}

Nematic liquid crystals are usually described  by the unit vector $\bn$, known as the director, with $\bn \equiv -\bn$ to obtain an orientation rather than a direction. In the continuum description the elastic Frank free energy~\cite{book:intro} is quadratic in gradients of $\bn$, including bend $\nabla\times\bn$ and splay $\nabla\cdot\bn$ terms in 2D. In the one elastic constant approximation (equal bend and splay contributions) the elastic energy simplifies to~\cite{book:intro}  
\begin{equation} \label{eq:fel}
{\cal F}_{el}=\frac K2\iint_\Omega dx\,dy\,|\nabla\bn|^2,
\end{equation}
where the integration is performed over the domain $\Omega$. In an infinite system the ground state corresponds to the uniform director field, $\bn=\const$. When  confined, liquid crystals reorient  to satisfy the boundary conditions, resulting in spatial variation of the director field, parametrized in Cartesian $(x,y)$ coordinates by (see Fig.~\ref{fig:charge}a)
\be\label{eq:bn}
\bn=\cos\theta(x,y)\,\bi_x+\sin\theta(x,y)\,\bi_y.
\ee
Minimizing the free energy~\eqref{eq:fel} leads to the Euler--Lagrange equation 
\be\label{eq:EL}
\p_{xx}\theta+\p_{yy}\theta=0,\quad\implies \quad \theta(z)=\sum_i{\Im}\big\{\log(z-z_i)^{q_i}\big\}, 
\ee
where  $z\equiv x+i y$. The equilibrium configurations described in Eq.~\eqref{eq:EL} correspond to a set of defects of strength (topological charge) $q_i$ located at positions $z_i$. Note that in the far field ($|z-z_i|\to \infty$) the director is uniform, which is a good approximation for systems with infinite size. The total topological charge $\sum_i q_i$ in~\eqref{eq:EL}, associated with the line field $\bn$ in 2D, is defined by enclosing  the defects at positions $z_i$ by an arbitrary contour $\gamma_1$ (see Fig.~\ref{fig:charge}b) and computing the following integral~\cite{book:intro,bowick:review}
\be\label{eq:qi}
\frac 1{2\pi}\oint_{\gamma_1}\!\!ds\,(\bn\times\p_s\bn)=\frac 1{2\pi}\oint_{\gamma_1}\!\!d\theta\mcm{=\sum_i q_i}.
\ee

The effects of confinement may be studied by adding an effective surface energy ${\cal F}_a$ favoring planar anchoring in the Rapini--Papoular form~\cite{book:intro}
\be\label{eq:fa}
{\cal F}_a=\frac {W_a}2\int_{\p\Omega}\!\!\ ds\,(\bn\cdot \bnu)^2,\qquad \bn\cdot\bnu=\cos(\theta-\vp)
\ee
where $W_a>0$ is the anchoring strength, $s$ is a curvilinear coordinate of the boundary $\p\Omega$ and $\bnu=\cos\vp(s)\,\bi_x+\sin\vp(s)\,\bi_y$ (see Fig.~\ref{fig:anchor}) is a unit normal to the boundary. The anchoring length $L_a=K/W_a$ is the length scale over which the director reorients to align tangentially with the boundary. Any simply connected domain $\Omega$ with boundary $\p\Omega$ is homeomorphic to a disc, whose Euler--Poincar\'e characteristic $\chi=1$~\cite{book:docarmo}, with
\be\label{eq:chi}
\chi=\frac 1{2\pi}\oint_{\p\Omega} ds\,(\bnu\times\p_s\bnu)=\frac 1{2\pi}\oint_{\p\Omega}d\vp.
\ee
We have chosen the counter-clockwise orientation of $\p\Omega$ as a positive one (Fig.~\ref{fig:charge}). The Euler--Poincar\'e characteristic of an annulus is zero ($\chi=0$). This can be shown by i)~integrating~\eqref{eq:chi} over two circles with opposite orientation, connected by a cut or ii)~triangulating an annulus and counting the number of vertices $V$, edges $E$ and faces $F$, yielding $\chi=V-E+F=0$.

%%%%%%%%%%%%%%%%%%%%%%%%%%%%%%%%%
%%% 	Figure 2 introduce charge
%%%%%%%%%%%%%%%%%%%%%%%%%%%%%%%%

If the characteristic size of our system $R$ is much smaller than the anchoring length,  $R\lesssim L_a$, the director remains uniform,  as shown in Fig.~\ref{fig:charge}a. Very strong anchoring $W_a$ ($R\gg L_a$), on the contrary, forces a preferred orientation of $\bn$ relative to the normal $\bnu$ (in the case of planar anchoring $\bn\perp\bnu$). Topological defects of charge $q_i$ then appear in the bulk, with  $\sum_i q_i=\chi$ (Poincar\'e--Hopf theorem~\cite{book:docarmo}). \mcm{On the other hand,  the most common situation in experiments corresponds to  $R\gtrsim L_a$. In this case surface anchoring and the energetic cost of bulk director deformations compete with each other and bulk defects can be pushed towards the boundary, resulting in  director deformations that satisfy the boundary conditions almost everywhere on $\p\Omega$, except at a finite number of points $z_j^\p$~\cite{carlson:1988}. The repulsive nature of the pairwise  defect--defect interactions favours maximal separation between same-sign defects~\cite{book:intro,bowick:review} while anchoring mediated interactions between topological defects have not been analyzed in detail before.

To characterize  defects localized on a 1D boundary $\p\Omega$, we introduce their topological charge $m_j$ } related to the angle deficit of the director field $\bn$, which rotates in the same (opposite) sense as the  interior contour $\gamma_2$ (see Fig.~\ref{fig:charge}b), yielding a positive (negative) charge $m_j= \Delta\theta_j/(2\pi)$. To establish the connection between the strengths $q_i$ and $m_j$ of bulk and surface defects, the topology of the system as described by $\chi$, and the anchoring $\bn\cdot\bnu$ we derive the charge conservation law,   following the approach outlined in~\cite{book:intro,volovik:1983} for 3D nematic systems. We choose a closed curve $\gamma_2$ going around the boundary defects $m_j$ at positions $z_j^\p$ (see Fig.~\ref{fig:charge}b). Since there are no special points of the director field $\bn$ between the curve $\gamma_1$, enclosing defects in the bulk, and $\gamma_2$, we get
\be\label{eq:curve}
2\pi \sum_i q_i\stackrel{\eqref{eq:qi}}{=}\oint_{\gamma_1}\!\!\!ds\,(\bn\times\p_s\bn)=\oint_{\gamma_2}\!\!\!ds\,(\bn\times\p_s\bn).
\ee
Decomposing the integral over the closed contour $\gamma_2$ into the contribution for the portion around the boundary defects, given by $-\sum_j m_j$, and the integral along piecewise regular curves, we find
\be\label{eq:gamma2}
\int_{\gamma_2\backslash\{z_j^\p\}} \kern-10pt ds\, \big[(\underbrace{n_\nu^2+n_\tau^2}_{1})\p_s\vp+\underbrace{n_\nu\p_sn_\tau-n_\tau \p_s n_\nu}_{\p_s\theta-\p_s\vp}\big]\bnu\times \btau.
\ee
The integral of the first term is $2\pi\chi$~\eqref{eq:chi}. The other terms can be computed using the parametrization  $\bn=n_\nu\,\bnu+n_\tau\,\btau$ in the local system of coordinates and the relations $\p_s\bnu\equiv \p_s\vp\,\btau$, $\p_s\btau=-\p_s\vp\,\bnu$. Replacing~\eqref{eq:gamma2} in~\eqref{eq:curve}, we get  the following charge conservation law 
\be\label{eq:conserv}
\sum_i q_i+\sum_j m_j=\chi +\frac 1{2\pi}\sum_j\int_{\gamma_2\backslash\{z_j^\p\}}\kern-10pt d(\theta-\vp). 
\ee
In Fig.~\ref{fig:disc} we show the transformation of a configuration of the director  field $\bn$ containing bulk defects   with $q_i=+2,+1$ to  configurations with boundary defects with $m_j=+\frac12,+\frac14$. By choosing an arbitrary curve $\gamma_2$ enclosing the defects, one can show that~\eqref{eq:conserv} holds in all the  cases considered in Fig.~\ref{fig:disc}. In fact the bulk and boundary defects are in the same topological class since the latter are obtained simply by pushing a bulk defect to the boundary. Here we restrict ourselves to smooth boundaries and thus $m\equiv q/2$.   

To find nematic configurations  minimizing the total free energy ${\cal F}_{el}+{\cal F}_a$ one needs to solve the Euler--Lagrange equation~\eqref{eq:EL} in the domain $\Omega$ with  the boundary condition, arising from the vanishing of the first variation of $\delta({\cal F}_{el}+{\cal F}_a)=0$ on $\p\Omega$, given by
\be\label{eq:bc}
K\bnu\cdot\nabla\theta-W_a\sin(\theta-\vp)\cos(\theta-\vp)=0.
\ee
Instead of solving this variational problem numerically, we propose a plausible  {\it ansatz} for the angle $\theta$ \mcm{that minimizes ${\cal F}_{el}$~\eqref{eq:EL}. We then seek  approximate solutions compatible with the one-, two- and three-fold symmetries  observed in experiments~\cite{thesis:2013} and compare their relative energies to determine the most favorable configuration. This allows us to treat the problem analytically.  We consider  configurations with different number of defects (up to 6) and topological charge $|m|\leqslant \frac 1 2$ (or $|q|\leqslant 1$), confined to a disc and an annular geometry. Since the boundaries of both disc and annulus have constant curvature, the position of defects correspond to the furthest separation along the boundary. The defect charge is, however, not known {\it a priori}.  Using this  approach, we compare the energetics of nematic liquid crystals confined to a disc or annulus and study the interplay between  the anchoring extrapolation length $L_a$, the system size $R$ and the core size of defects $\ve$ in controlling the lowest energy configurations.}

%%%%%%%%%%%%%%%%%%%%%%%%%%%%%%%%%%%%%%%%%%%%
\section*{Nematic confined to a disc}
%%%%%%%%%%%%%%%%%%%%%%%%%%%%%%%%%%%%%%%%%%%

\begin{figure}[th]
\centering
(a)~$q=+2$\hskip1cm (b)~$q=+1$\hskip1cm (c)~$m=+1/2$
\vskip1ex
\includegraphics[height=2.6cm]{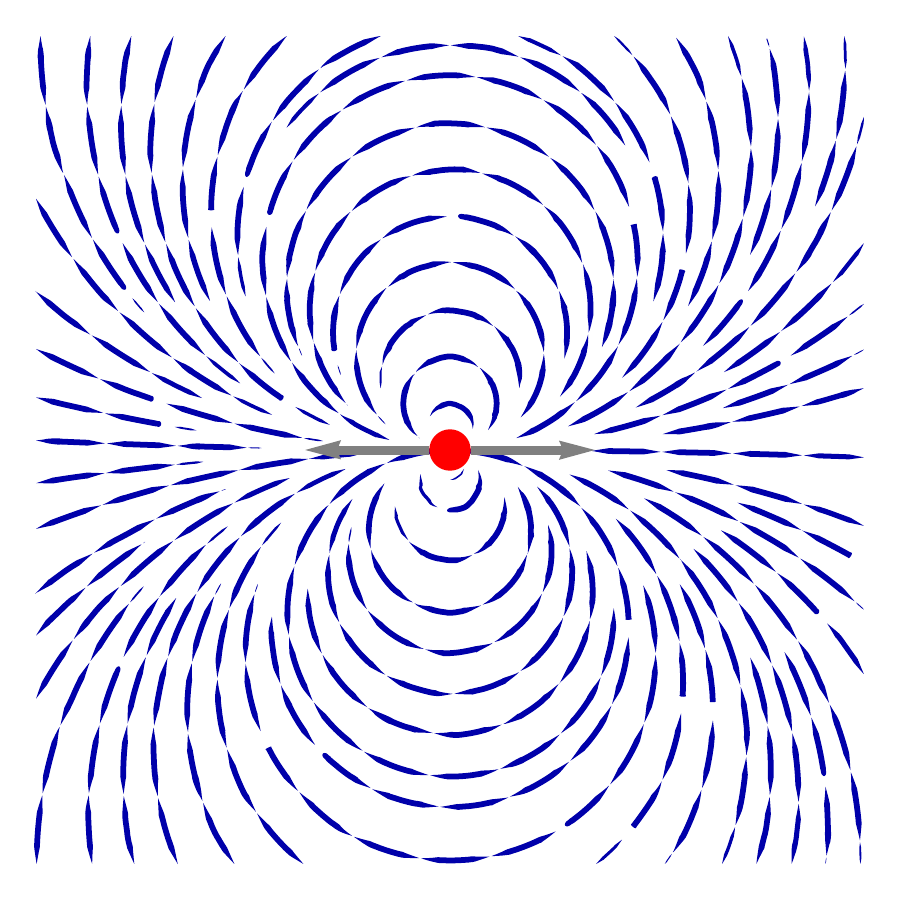}
\hfil
%\raisebox{20mm}{\kern-10pt\Large$\Rightarrow$\kern10pt}
\includegraphics[height=2.6cm]{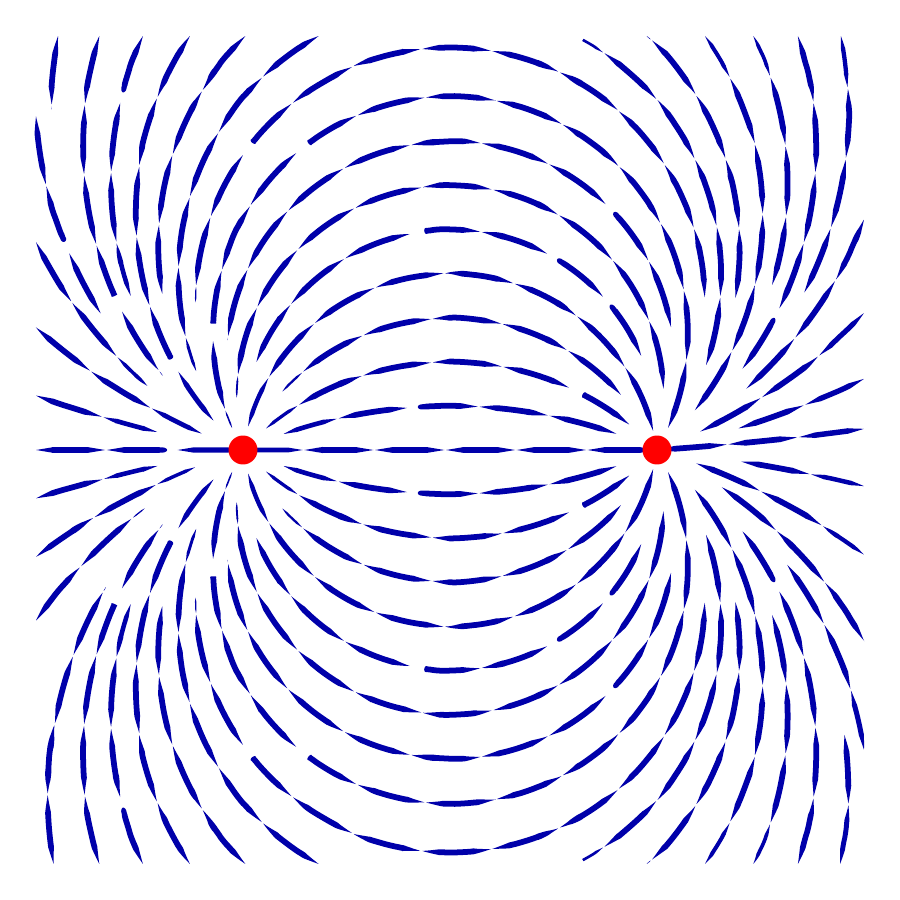}
\hfil
%\raisebox{20mm}{\kern-10pt\Large$\Rightarrow$\kern10pt}
\includegraphics[height=2.6cm]{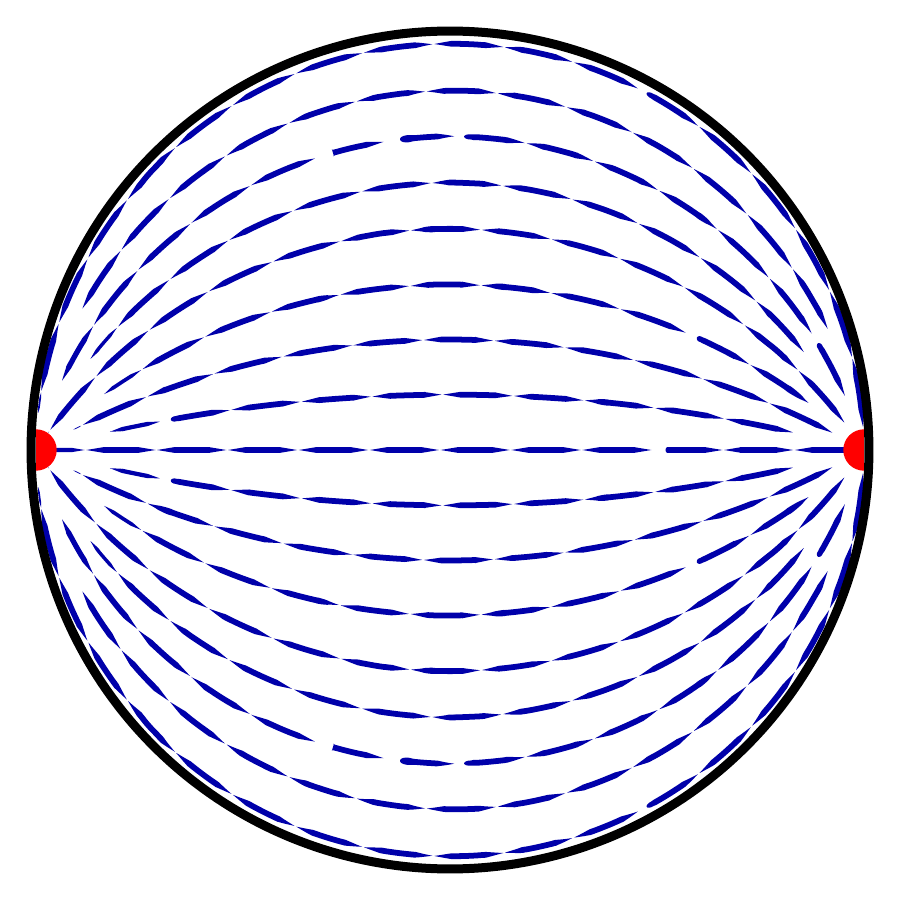}
\vskip1ex
(d)~$q=+1$\hskip1cm (e)~$q=+\frac12$\hskip1cm (f)~$m=+ 1/4$
\vskip1ex
\includegraphics[height=2.6cm]{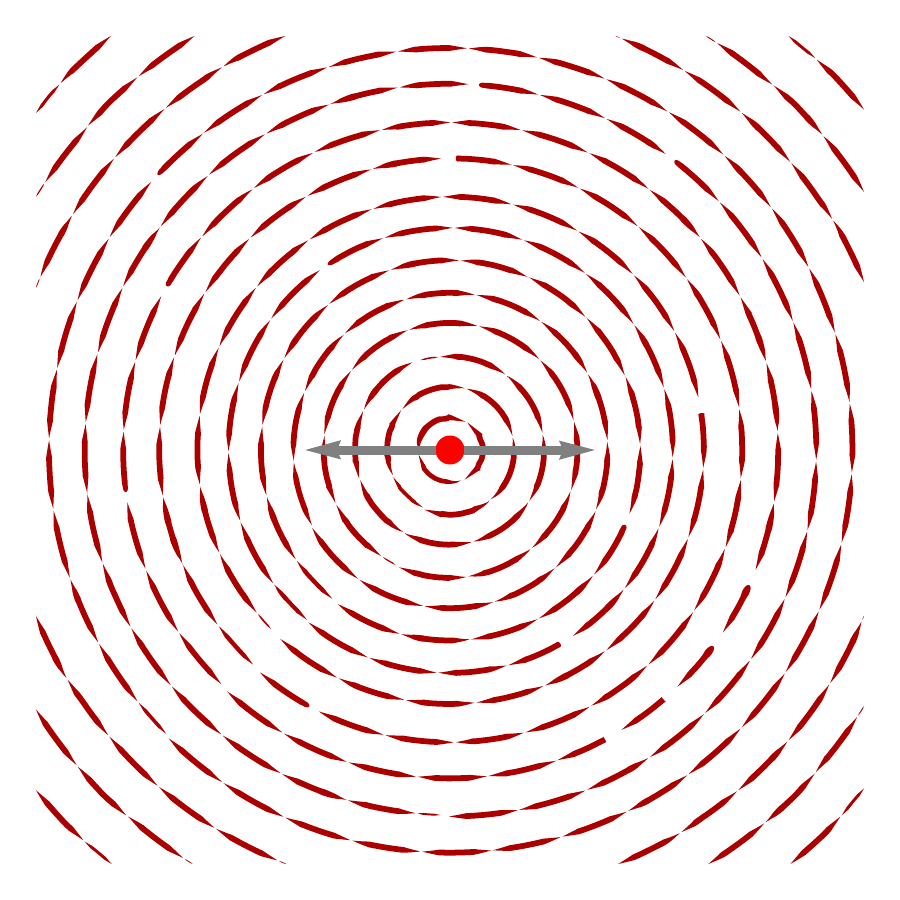}
\hfil
%\raisebox{20mm}{\kern-10pt\Large$\Rightarrow$\kern10pt}
\includegraphics[height=2.6cm]{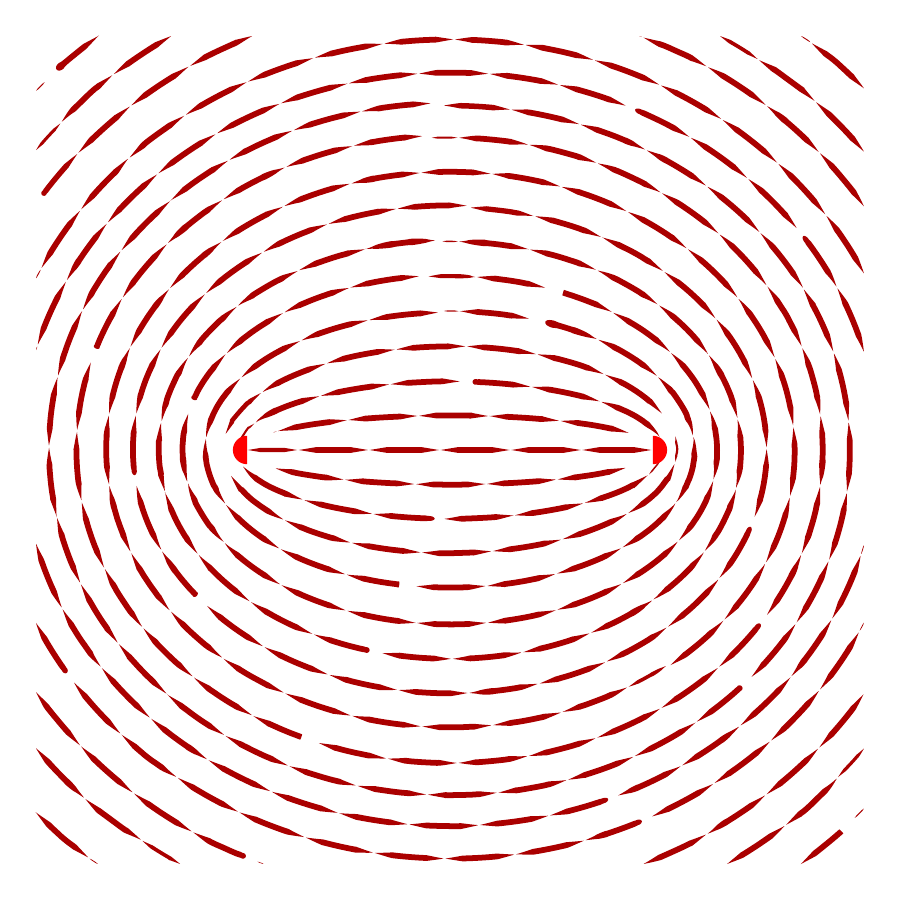}
\hfil
%\raisebox{20mm}{\kern-10pt\Large$\Rightarrow$\kern10pt}
\includegraphics[height=2.6cm]{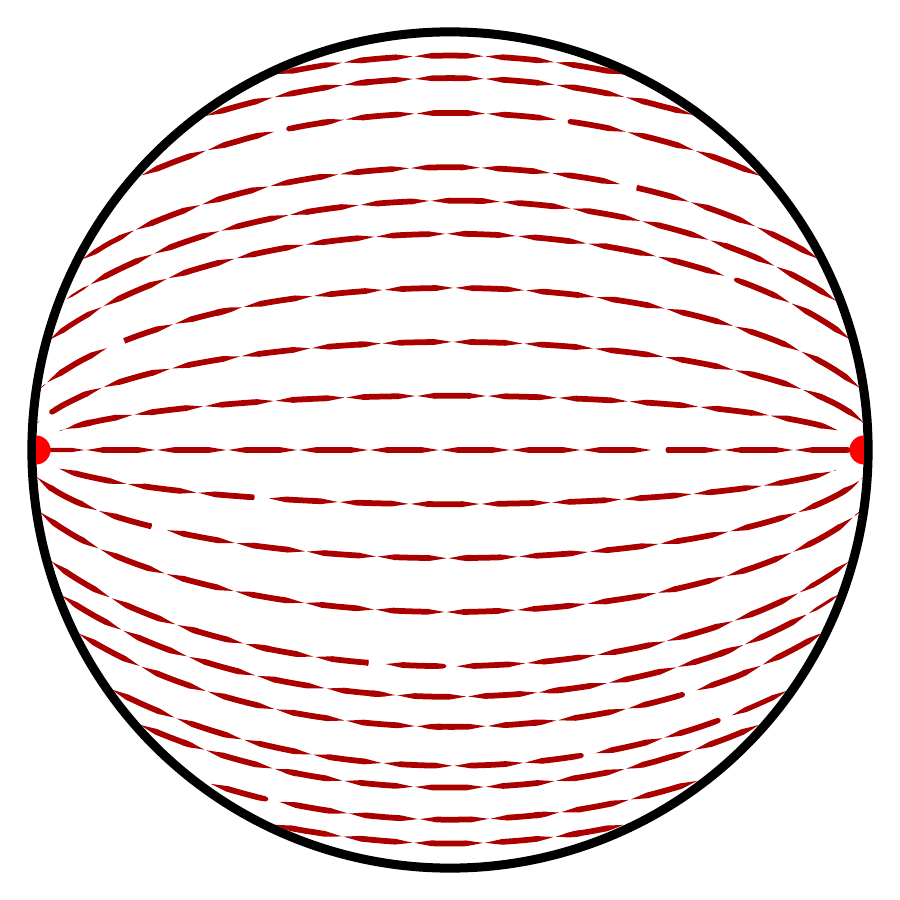}
\caption{\label{fig:disc} (a)--(c) Shows the transformation of a bulk defect of charge $+2$  into two $+1$ defects, which subsequently approach the boundary of the circular region, where they can be classified as charge $m=+\frac12$ boundary defects. (d)--(f)~Similarly, a charge $q_i=+1$ defect is split into two $q_i=+\frac12$ defects which in turn yield two charge $m=+\frac14$ boundary defects. This process illustrates the construction of our {\it ansatz}. The parametrization of the director field configuration of (c) and (f) is given by~\eqref{eq:thetac} and \eqref{eq:thetaf}, respectively.}
\end{figure}

The vector field $\bn$~\eqref{eq:bn} shown in Fig.~\ref{fig:disc}b,e without confinement can be written explicitly as 
\begin{align}\label{eq:thetac}
\theta^{(+\frac 12)}(z)&=\Im \big\{\log(z^2-R^2)\big\}, \quad z\equiv r e^{i\vp},\\
\theta^{(+\frac 14)}(z)&=\frac \pi2+\Im\big\{ \log\sqrt{z^2-R^2}\big\}. \label{eq:thetaf}
\end{align}
It accounts for the pairs of topological defects at positions $z_i=\pm R$ with charges $q_i=+1$ and $q_i=+\frac12$ and satisfies the Euler--Lagrange equation~\eqref{eq:EL}. Confining the director fields given by Eqs.~\eqref{eq:thetac}, \eqref{eq:thetaf} to a disc of radius $R$  (Fig.~\ref{fig:disc}c,f) leads to  boundary defects characterized by  $m=+\frac12$ and $m=+\frac14$, respectively, while the director $\bn$ remains almost uniform in the bulk. \mcm{According to Eq.~\eqref{eq:conserv} we expect no anchoring contribution for a pair of $m =+\frac1 2$ where the boundary condition is not satisfied just at the defect core and  the total charge equals  $\chi$, while for a pair with  $m=+\frac 1 4$  the deviation from the preferred anchoring orientation extends over a finite portion of the boundary, yielding a non-zero anchoring contribution.}

%%%%%%%%%%%%%%%%%%%%%%%%%%%%%%%%%
%%% 	Figure disc
%%%%%%%%%%%%%%%%%%%%%%%%%%%%%%%%

Below we quantify the energetics of the nematic configurations~(Fig.~\ref{fig:disc}c,f) and analyze the  stability of \mcm{configurations with boundary defects relative}  to the defect free, uniform state. \mcm{We assume that the size $\ve$ of the defect core  satisfies $\ve\ll R$  and that the associated core energy is much smaller than the elastic energy in the bulk and thus can be neglected.} First we compute the total free energy of configuration with $m=+\frac 14$. Substituting the nematic angle $\theta^{(+\frac14)}(z)$ given in Eq.~\eqref{eq:thetaf}  into the anchoring and elastic free energy ${\cal F}_a$ and ${\cal F}_{el}$ given in Eqs.~\eqref{eq:fa} and ~\eqref{eq:fel}  and integrating  over the polar coordinate $\vp$, we find 
\begin{gather}
{\cal F}_a^{(+\frac 14)}=\frac {W_aR}2\int_0^{2\pi}\!\! d\vp\cos^2(\theta^{(+\frac 14)}\!\!-\vp)= \frac{W_a R}2(\pi-2),\\
{\cal F}_{el}^{(+\frac 14)}= \frac K2 \int_0^{2\pi}\!\!d\vp\int_0^R \frac{dr\,r^3}{R^4+r^4-2r^2R^2\cos(2\varphi)}\simeq\notag\\\simeq \frac K2\bigg\{\frac \pi 2 \log \frac R {4\ve} +\frac \ve R\bigg(\frac{3\pi}4+\cot\bigg(\frac \ve R\bigg)\bigg)+O\bigg(\frac\ve R\bigg)^2\bigg\},\label{eq:f14}
\end{gather}
where we have truncated the converging power series in $\ve/R$. Including the next order corrections may be appropriate for systems with relatively large defect core size $\ve$. In conventional nematics $\ve$ is of the order of nanometers, which is the characteristic scale of the constituent  molecules. In Ref.~\cite{aarts:2014}, the authors assumed $\ve=0.88~\mu$m for {\it fd}-virus and estimated the anchoring extrapolation length $L_a\simeq 1.4~\mu$m.

Nematic configurations with a pair of boundary defects of charge  $m=+\frac12,$ in a disc~(Fig.~\ref{fig:disc}c) have zero anchoring energy, except for a small region $O(\ve)$ around the defects  where the anchoring condition is not satisfied. Their elastic energy is, however, four times higher than ${\cal F}_{el}^{(+\frac14)}$~\eqref{eq:f14} since it is proportional to the square of the charge. Notice also that the structure with a bulk defect $q_i=+1$~(Fig.~\ref{fig:disc}d) confined to a disc is unstable with respect to configuration~(c), since ${\cal F}_d-{\cal F}_c\propto 2\pi K\log2$, neglecting the contribution from the defect core. The fact that no bulk defects were observed in experiments with  {\it fd}-virus~\cite{thesis:2013} suggests that these systems are characterized by a finite anchoring strength~$W_a$. 
Finally, pushing  defects of higher charge, such as $q=+2$~(Fig.~\ref{fig:disc}a), to the boundary of a disc~\cite{lubensky:1999} gives higher elastic energy $\simeq2 \pi K \log (2R/\ve)$ compared to the configurations shown in Figs.~\ref{fig:disc}(c,d).

%%%%%%%%%%%%%%%%%%%%%%%%%%%%%%%%%
%%% 	Figure phase diagram disc 
%%%%%%%%%%%%%%%%%%%%%%%%%%%%%%%%^{(+\frac12)}

\begin{figure}[t]
\centering
\includegraphics[width=0.65\linewidth]{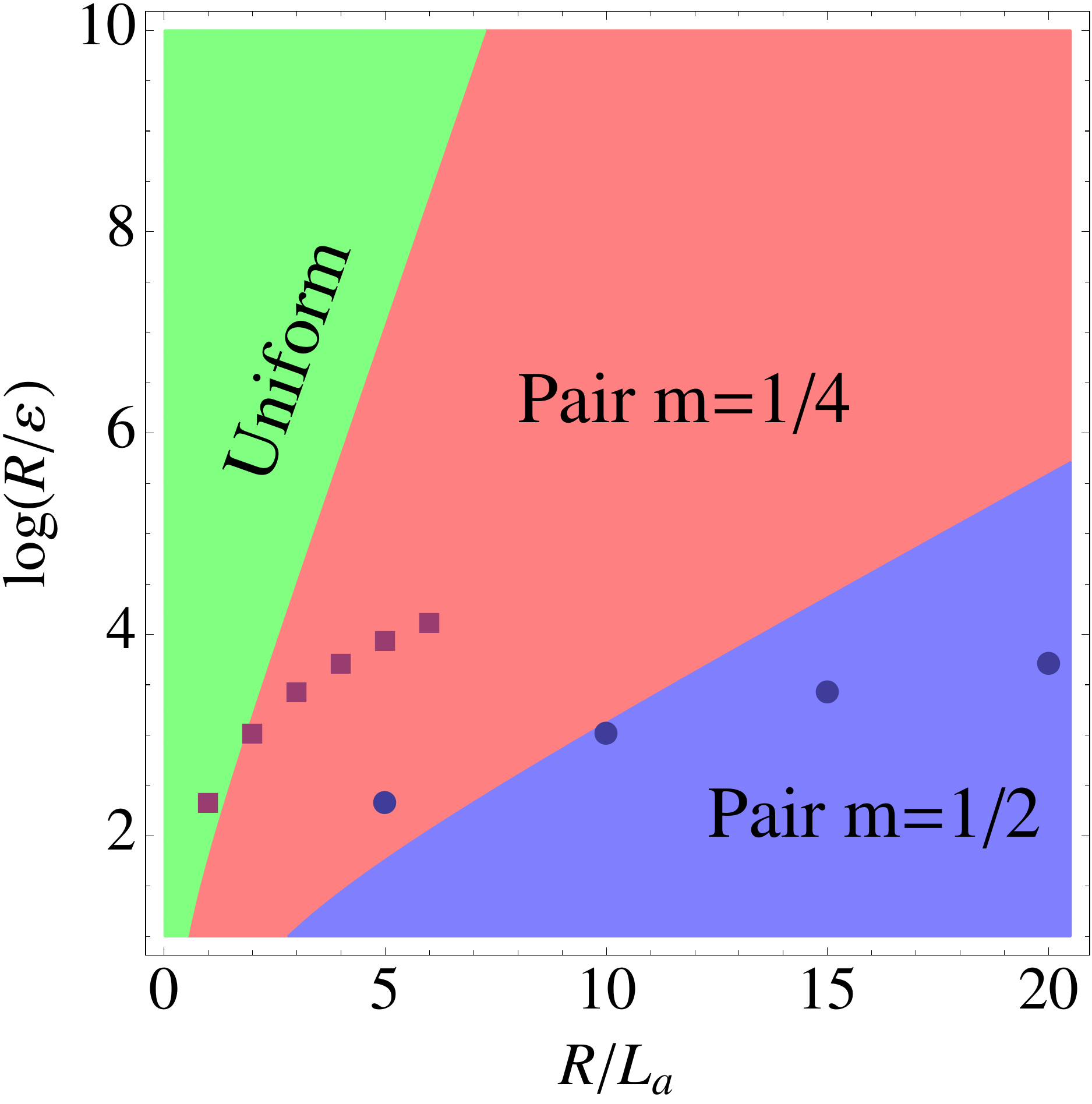}
\caption{\label{fig:dphase}\mcmred{Phase diagram of nematic film confined to a disc of radius $R$ showing the regions of parameters corresponding to the energy minimum for three configurations: a uniform state (green) and the two defective configurations with charges $m=+\frac12$ (blue) and $m=+\frac14$ (pink) shown in Fig.~\ref{fig:disc}c,f. The horizontal axis is the dimensionless parameter $R/L_a=R W_a/K$; the vertical axis is the ratio of the radius $R$ to the defect core $\ve$~\eqref{eq:f14}. The circles show the location of the crossover between the $m=\frac1 4$ defect configuration and the uniform state for $L_a=1~\mu$m and various values of the radius of the disc ($R=5,10,15\ldots~\mu$m from left to right).
The squares denote the corresponding crossover between the two defective configurations for $L_a=5~\mu$m In both cases we have  chosen $\ve=0.5~\mu$m.}}
\end{figure}

To quantify the role of anchoring $L_a=K/W_a$ and system size $R$ we compute the phase diagram. In Fig.~\ref{fig:dphase} we compare three nematic states confined to a disc of size $R$: a uniform state $\bn=\const$, and two nematic configurations  with a pair of boundary defects separated by $2R$ and charge $m=+\frac1 2$ and $ m=+\frac14$, respectively (see Fig.~\ref{fig:disc}c,f), \eqref{eq:thetaf}. \mcm{The relative energy of these configurations is controlled by the interplay of the defect core size $\ve$,  the anchoring strength $L_a$ and the radius $R$ of the disc. Note that the core energy for a pair of defects is of the order of $\pi K q^2$~\cite{book:intro}, independently of $\epsilon$, and accounting for this contribution will lead to a slight shift of the coexistence curves in the phase diagram without changing the basic picture. The uniform state is stable for small systems ($R$) and weak anchoring (large $L_a\gtrsim R$). A small value of $\ve$ increases the region of stability of the uniform state. For stronger anchoring (or smaller system size) the lowest energy state is one with a pair of boundary defects. The  $m=+\frac12$ defects are favored for large values of the core size $\ve$ while the configuration with $m=+\frac14$ is preferred for smaller core sizes. The main result is that by tuning the system-size $R$ one can drive a transition between a uniform state and a defective configuration selected by the minimization of the anchoring and elastic energies.  The transition is controlled by two dimensionless length scales $R/\ve$ and $R/L_a$, where $\ve$ and $L_a$ are determined by intrinsic physical and chemical properties of the system.}

Next, we compare our analytic predictions with experimental data for the {\it fd}-virus~\cite{thesis:2013} confined to a disc. \mcm{For weak anchoring ($L_a=5~\mu$m) we find that increasing the radius $R=5,10,15\ldots~\mu$m for a chosen core size $\ve=0.5~\mu$m yields a  transition from a uniform to a `defective' state with $m=\frac 14$ (squares in Fig.~\ref{fig:dphase}). For strong anchoring ($L_a=1~\mu$m) the transition is between the two defective  configurations  (circles in Fig.~\ref{fig:dphase}).}  
\mcm{In both cases the critical system size where the transition occurs is in the range $R\simeq 5-10~\mu$m, which is compatible with experiments~\cite{thesis:2013}. On the other hand, since the defect charge associated with reorientation of the $fd$ virus was not extracted from the experimental data~\cite{thesis:2013}, we cannot determine the value of the anchoring extrapolation length $L_a$ based   solely on the results shown in Fig.~\ref{fig:dphase}.}

In the following we adapt simplified analytical model to a viral nematics confined to an annular geometry. Our goal is to gain insight into the mechanism responsible for the selection of the symmetry of the boundary defect arrangement  (one-, two- or three-fold, as seen in experiments~\cite{thesis:2013}), as well as their topological charge, and estimate the value of the anchoring strength $L_a$ within our approximations.

%%%%%%%%%%%%%%%%%
\section*{Nematic in annular geometry}
%%%%%%%%%%%%%%%%%%

%%%%%%%%%%%%%%%%%%%%%%%%%%%%%%%%%
%%% 	Figure anchoring + coordinates
%%%%%%%%%%%%%%%%%%%%%%%%%%%%%%%%
\begin{figure}[tb]
\centering
\includegraphics[width=0.65\linewidth]{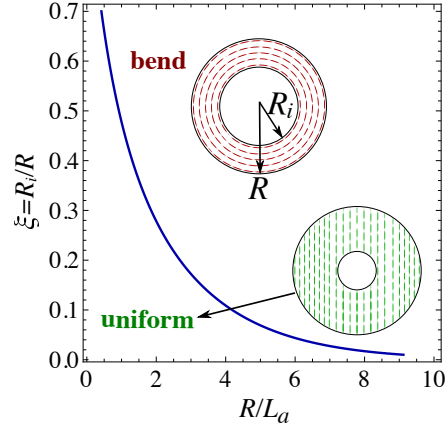}
\caption{\label{fig:anchor} The coexistence curve of the bend state with $\bn=-\sin\vp\,\bi_x+\cos\vp\,\bi_y$ and the uniform state with $\bn=\bi_y$ in an annulus of inner radius $R_i$ and outer radius $R$ (see Fig.~\ref{fig:charge} for the definition of the coordinate system). The horizontal axis is the system size $R$ scaled by the anchoring  extrapolation length $L_a=K/W_a$, the vertical axis is the ratio $\xi=R_i/R$. The bend state (${\cal F}_{el}>0$~\eqref{eq:fel}) is energetically favored above the curve, while the uniform state (${\cal F}_a>0$~\eqref{eq:fa}) has lower free energy in the region below the curve.}
\end{figure}

Now we consider a nematic liquid crystal confined to an annulus of inner radius $R_i$ and  outer radius $R$. This geometry does not require the presence of topological defects since the Euler characteristic of the annulus is  $\chi=0$. In this case the defect-free ground state shown in the top inset of Fig.~\ref{fig:anchor} satisfies the tangential boundary condition. We will refer to this structure as the bend configuration. In Fig.~\ref{fig:anchor} we compare the total free energy  ${\cal F}_{\rm tot}={\cal F}_{\rm el}+{\cal F}_{a}$ of a uniform state  and a bend state as a function of $R/L_a$ and $\xi=R_i/R\in(0:0.7]$. For small radius $R$  or weak anchoring (large $L_a$) a uniform state with zero elastic energy and  anchoring energy ${\cal F}_{a}=W_a\pi R(1+\xi)/2$ is energetically favored, consistent with the phase diagram in~Fig.~\ref{fig:disc}. For large $R/L_a$ a bend state with no anchoring contribution and elastic energy ${\cal F}_{el}=-K\pi\log\xi$ becomes energetically favorable. According to experiments with {\it fd}-virus~\cite{thesis:2013}, the transition between these two states occurs at $R\simeq5~\mu$m and $\xi\simeq 0.3-0.5$. Using this value, we estimate a value $L_a\simeq2.7-5.5~\mu$m, which is of the same order as  $L_a\simeq1.4~\mu$m obtained for {\it fd}-virus in rectangular geometries~\cite{aarts:2014} with the same physical properties of the boundary.

%%%%%%%%%%%%%%%%%%%%%
%%%% 	Defects within annular geometry
%%%%%%%%%%%%%%%%%%%%%
\begin{figure}[thb]
\centering
\raisebox{25mm}{$\boxed{1}$\kern-10pt}\includegraphics[height=2.7cm]{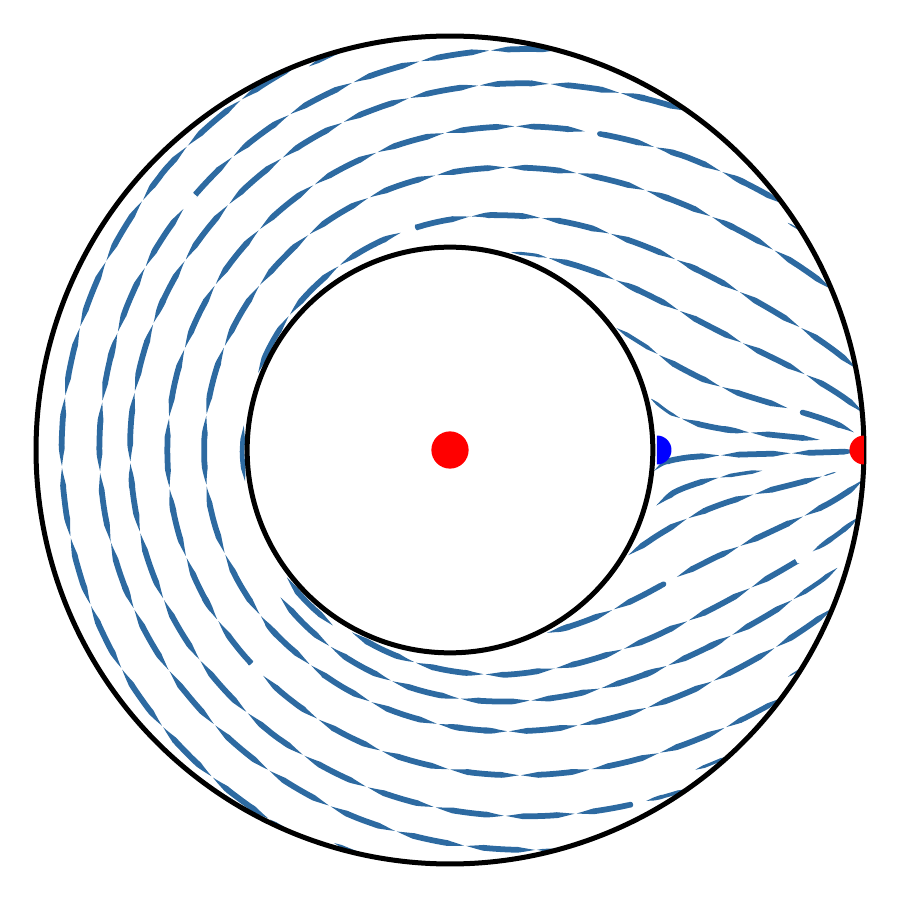}
\hfil
\raisebox{25mm}{$\boxed{2}$\kern-10pt}\includegraphics[height=2.7cm]{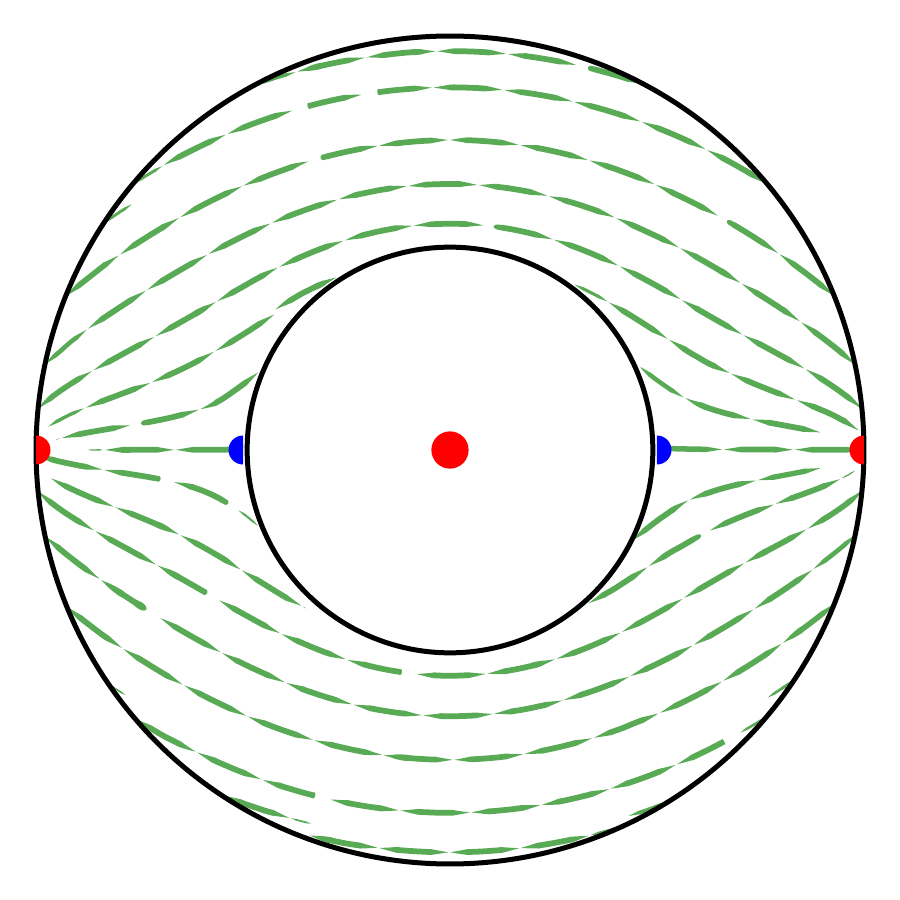}
\hfil
\raisebox{25mm}{$\boxed{3}$\kern-10pt}\includegraphics[height=2.7cm]{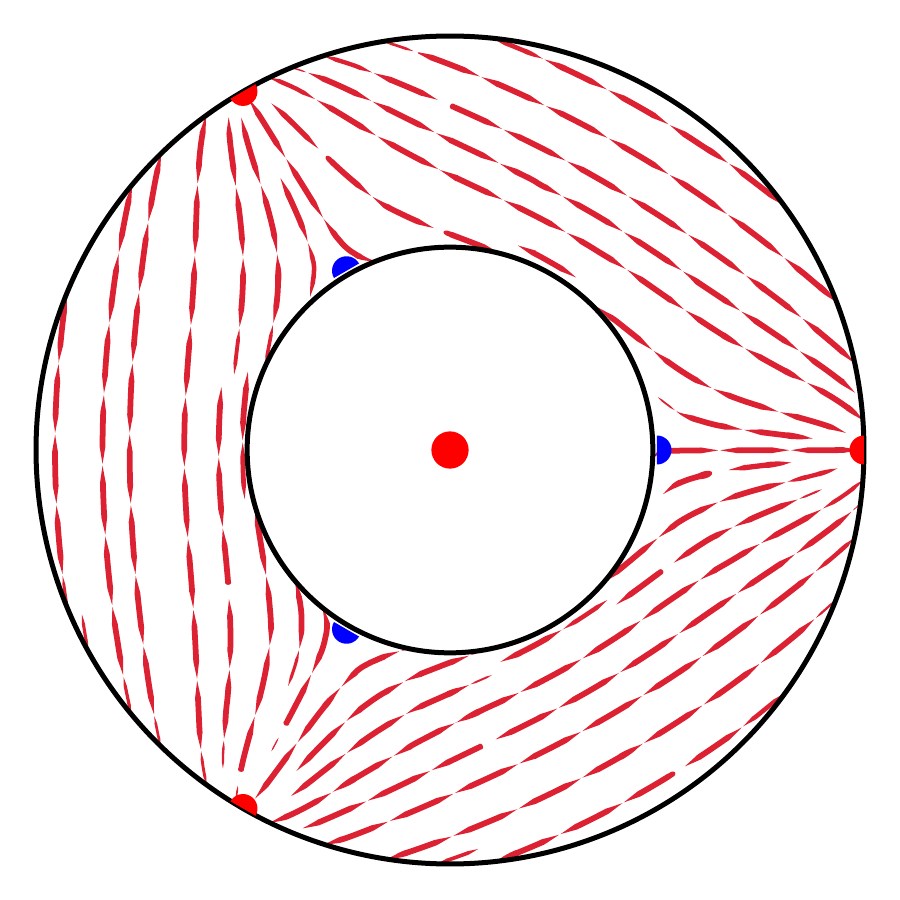}
\vskip1ex
\raisebox{25mm}{$\boxed{4}$\kern-10pt}\includegraphics[height=2.7cm]{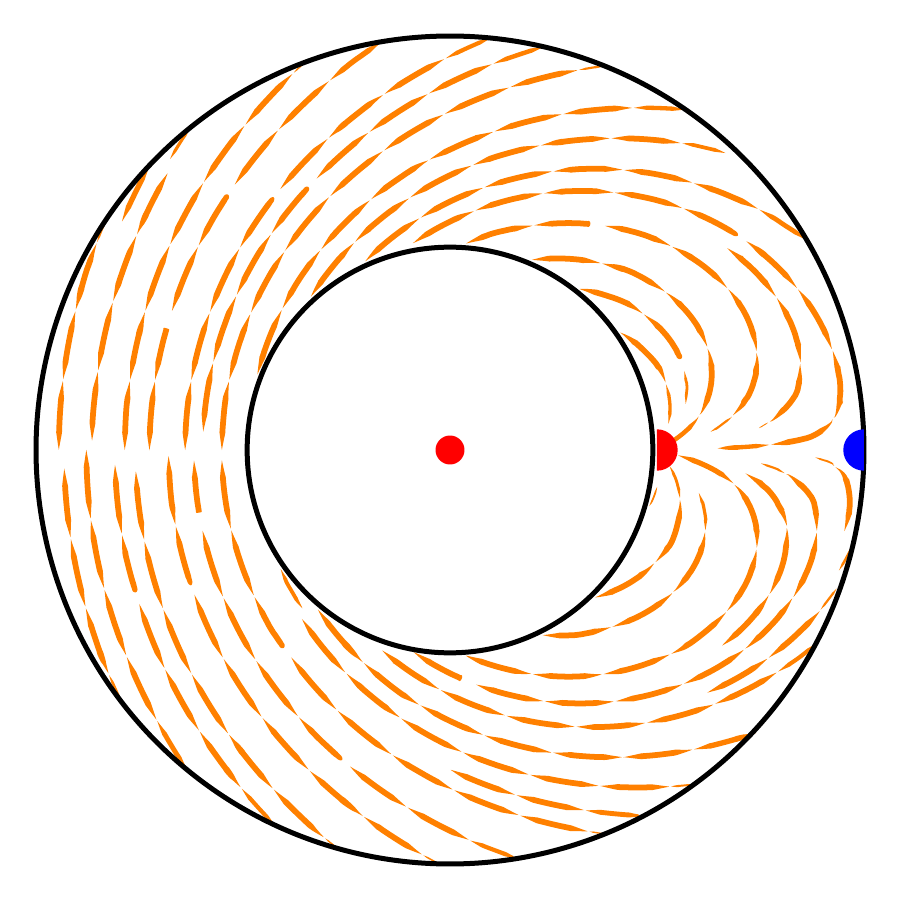}
\hfil
\raisebox{25mm}{$\boxed{5}$\kern-10pt}\includegraphics[height=2.7cm]{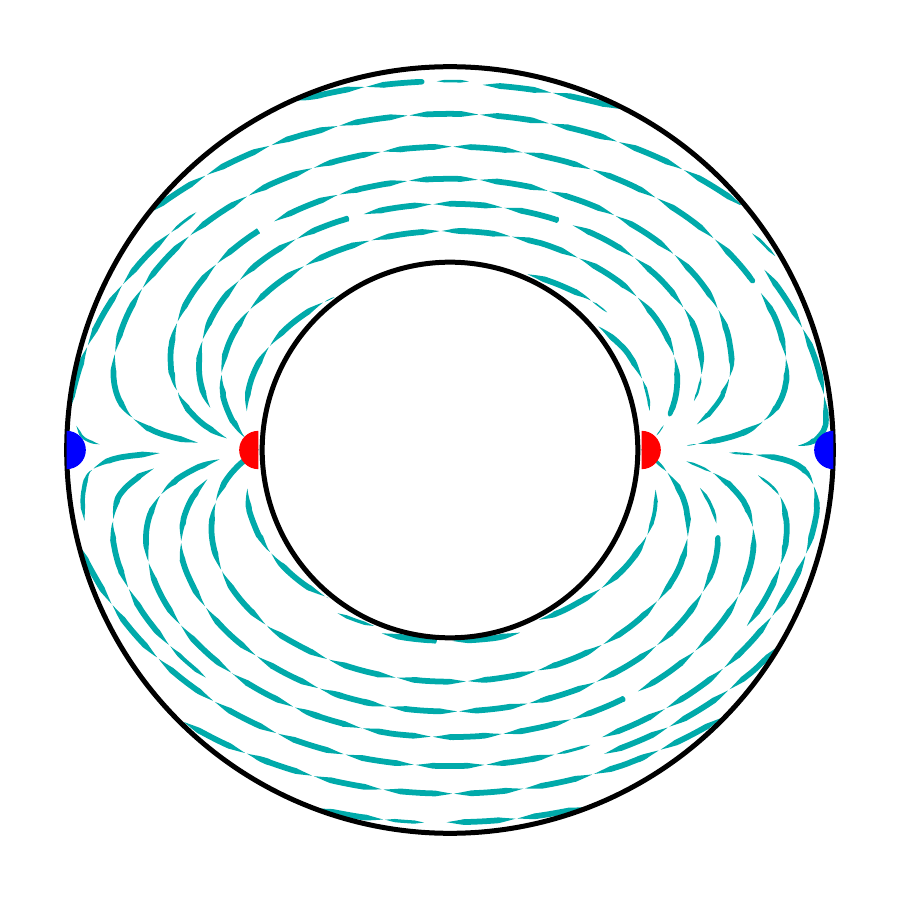}
\hfil
\raisebox{25mm}{$\boxed{6}$\kern-10pt}\includegraphics[height=2.7cm]{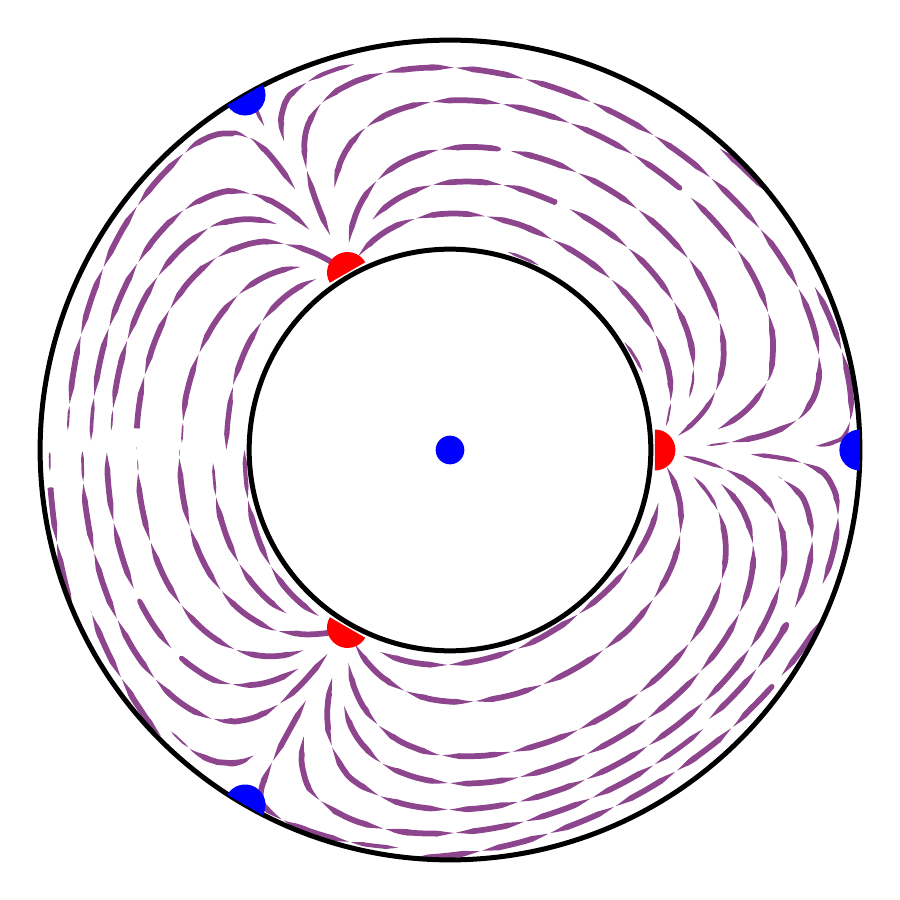}
\vskip1ex
\raisebox{25mm}{$\boxed{7}$\kern-10pt}\includegraphics[height=2.7cm]{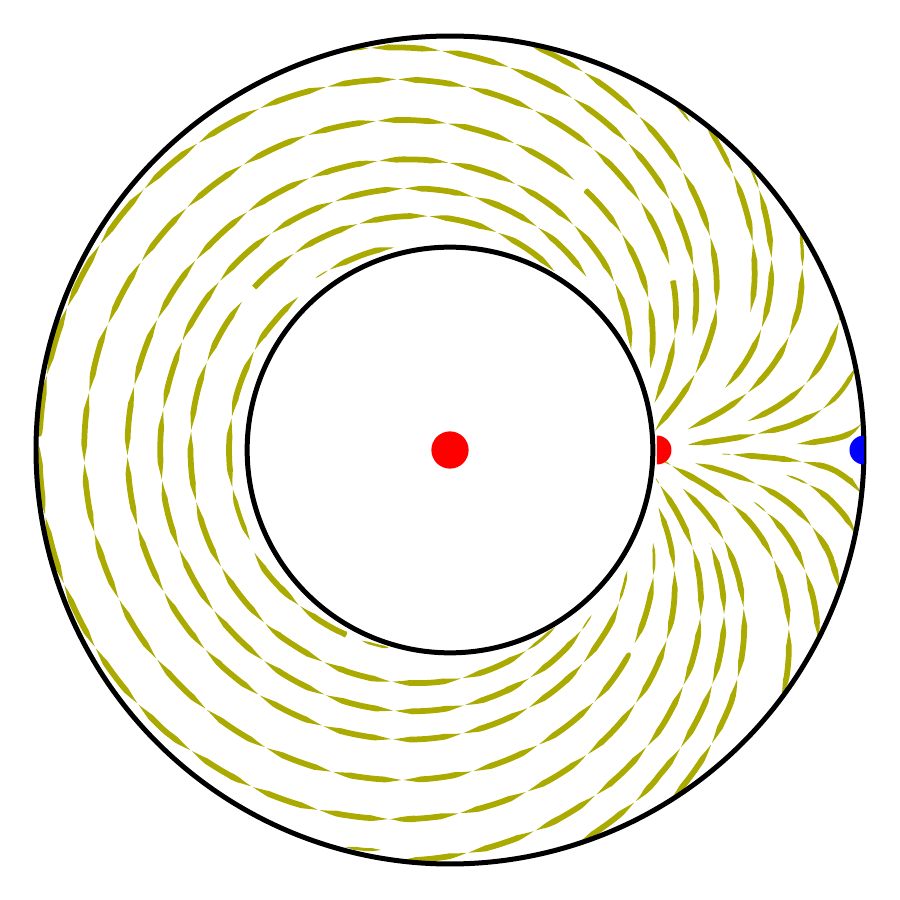}
\hfil
\raisebox{25mm}{$\boxed{8}$\kern-10pt}\includegraphics[height=2.7cm]{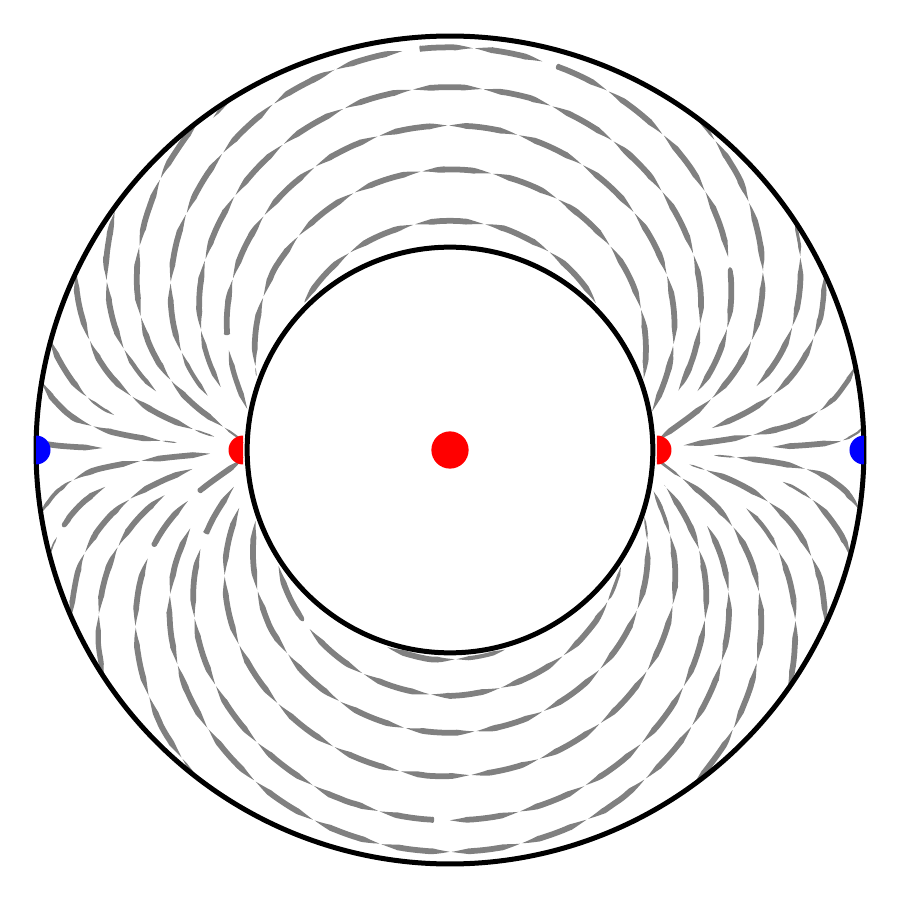}
\hfil
\raisebox{25mm}{$\boxed{9}$\kern-10pt}\includegraphics[height=2.7cm]{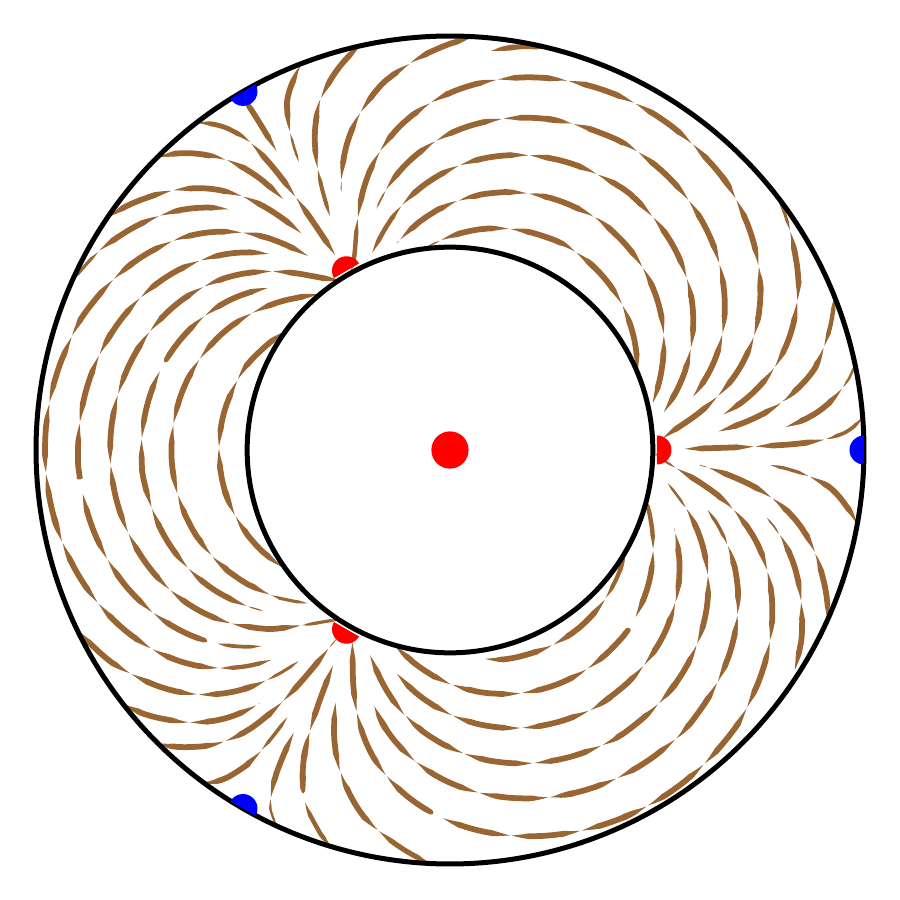}
\caption{\label{fig:annul}  Nematic textures confined to annular geometry. The size of dots describes the magnitude of the defect charge:  $m = \frac14$ ($\boxed{1}$--$\boxed{3}$ and $\boxed{7}$--$\boxed{9}$), $m=\frac12$ ($\boxed{4}$--$\boxed{6}$); the color corresponds to the sign, with  red denoting positive defects ($m>0$) and blue negative ones ($m<0$). The textures $\boxed{7}$--$\boxed{9}$ have the same anchoring energy but higher elastic energy than $\boxed{1}$--$\boxed{3}$, hence are not included in the stability phase diagram shown in Fig.~\ref{fig:aphase}.}
\end{figure}

%%%%%%%%%%%%%%%%%%%%%
%%%% 	Table with energy
%%%%%%%%%%%%%%%%%%%%%
\begin{table}[tb]
\caption{\label{tab:Ea}The total anchoring, ${\cal F}_a$, and elastic, ${\cal F}_{el}$, contributions to the free energy (integrated over the whole system) calculated \mcm{to lowest order in $\ve/R$} for the configurations $\boxed 1$--$\boxed 3$   (Fig.~\ref{fig:annul}). The director $\bn$ is parametrized by the angle $\theta=\frac \pi 2+\arg\big[\frac{z^\beta-R^\beta}{z^\beta-R_i^\beta}\big] $, where $\beta=1,2,3$ for the first, second and third row, respectively, and $\xi=R_i/R$.}
\centering
\medskip
{%\def\arraystretch{1.1}
\def\tabcolsep{5pt}
\def\extrarowheight{7pt}
%\begin{tabular}{m{10pt}@{}c@{}c}\hline %\hsize}{@{\extracolsep{\fill}}
\begin{tabular}{ccc}\hline %\hsize}{@{\extracolsep{\fill}}
& ${\cal F}_a$, $W_a R(1+\xi)/2$ &${\cal F}_{el}$, $K/2$ \\
\hline
\hline
$\boxed1$ & $\frac{2(1+\xi)(\frac\pi2-2 \arctan{\sqrt{\xi}})-\pi (1-\sqrt\xi)^2}{2\sqrt{\xi}}$
& $\frac \pi 2\log\big(\frac R\ve \cdot \frac{1-\xi^2}{2\sqrt\xi}\big)$\\[1ex]
%\hline
$\boxed2$ & $\frac{\pi  \xi -2(1+\xi^2)\arctan\xi}\xi$
& $ \pi\log\big(\frac R\ve \cdot \frac{\sqrt \xi(1-\xi^4)}4\big)$\\[1ex]
%\hline
$\boxed3$ &$\frac{(1+\xi^3)(\pi -4\arctan\xi^{3/2})-\pi(1-\xi ^{3/2})^2}{2\xi^{3/2}}$
& $ \frac{3\pi}2\log\big(\frac R\ve \cdot \frac{ \xi^{1/6}(1-\xi^6)}6\big)$\\[1ex]
\hline
\end{tabular}
}%
\end{table}

%%%%%%%%%%%%%%%%%%%%%%%%%%%%%%%%%
%%% 	Phase diagram annulus 
%%%%%%%%%%%%%%%%%%%%%%%%%%%%%%%%

\begin{figure}[tb]
\centering
\raisebox{35mm}{(a) \kern10pt}\includegraphics[width=0.65\linewidth]{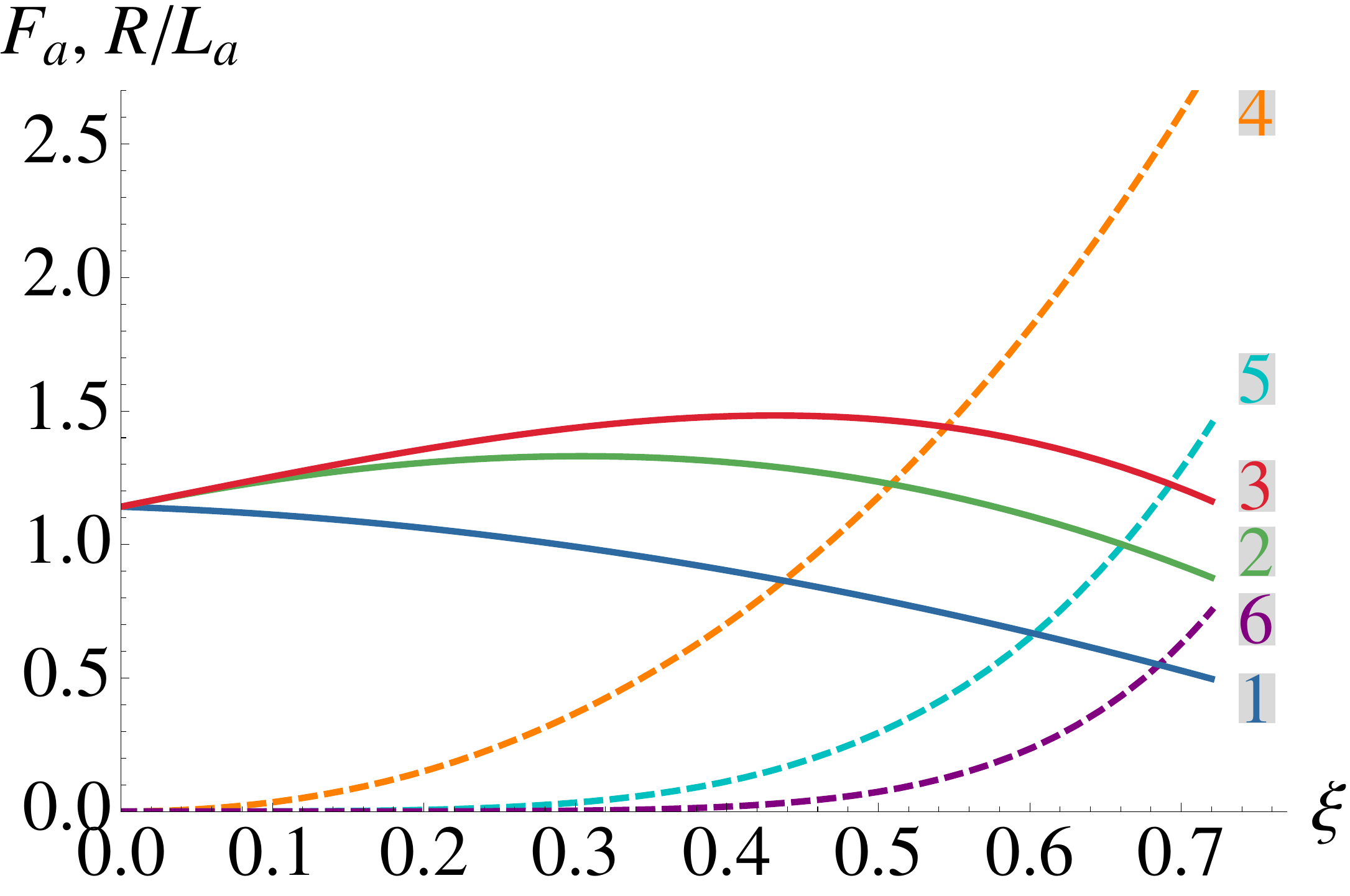}
\vskip3ex
\raisebox{43mm}{(b)\kern-5mm}\includegraphics[width=0.47\linewidth]{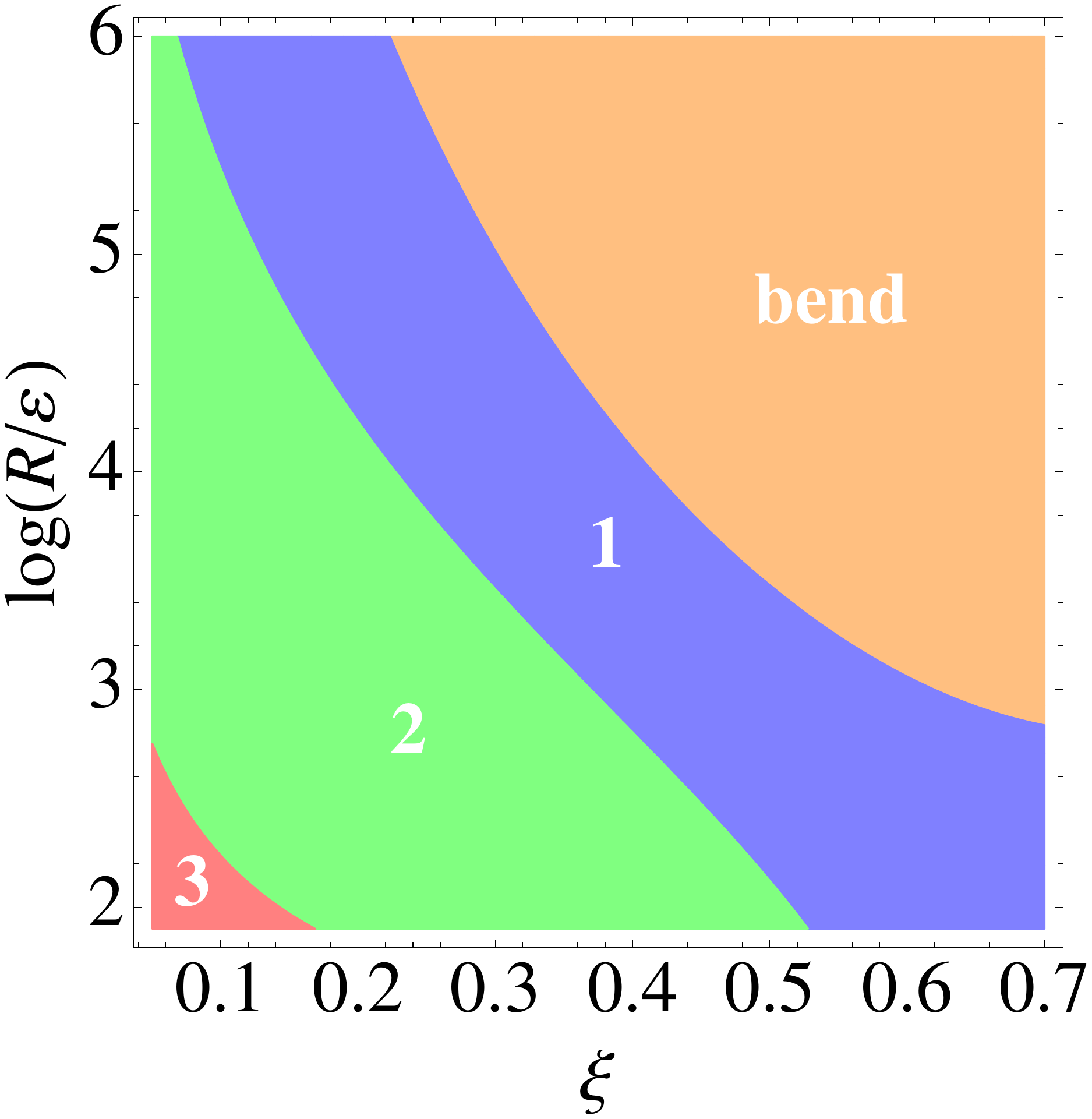}
\hfil
\raisebox{43mm}{(c)\kern-5mm}\includegraphics[width=0.47\linewidth]{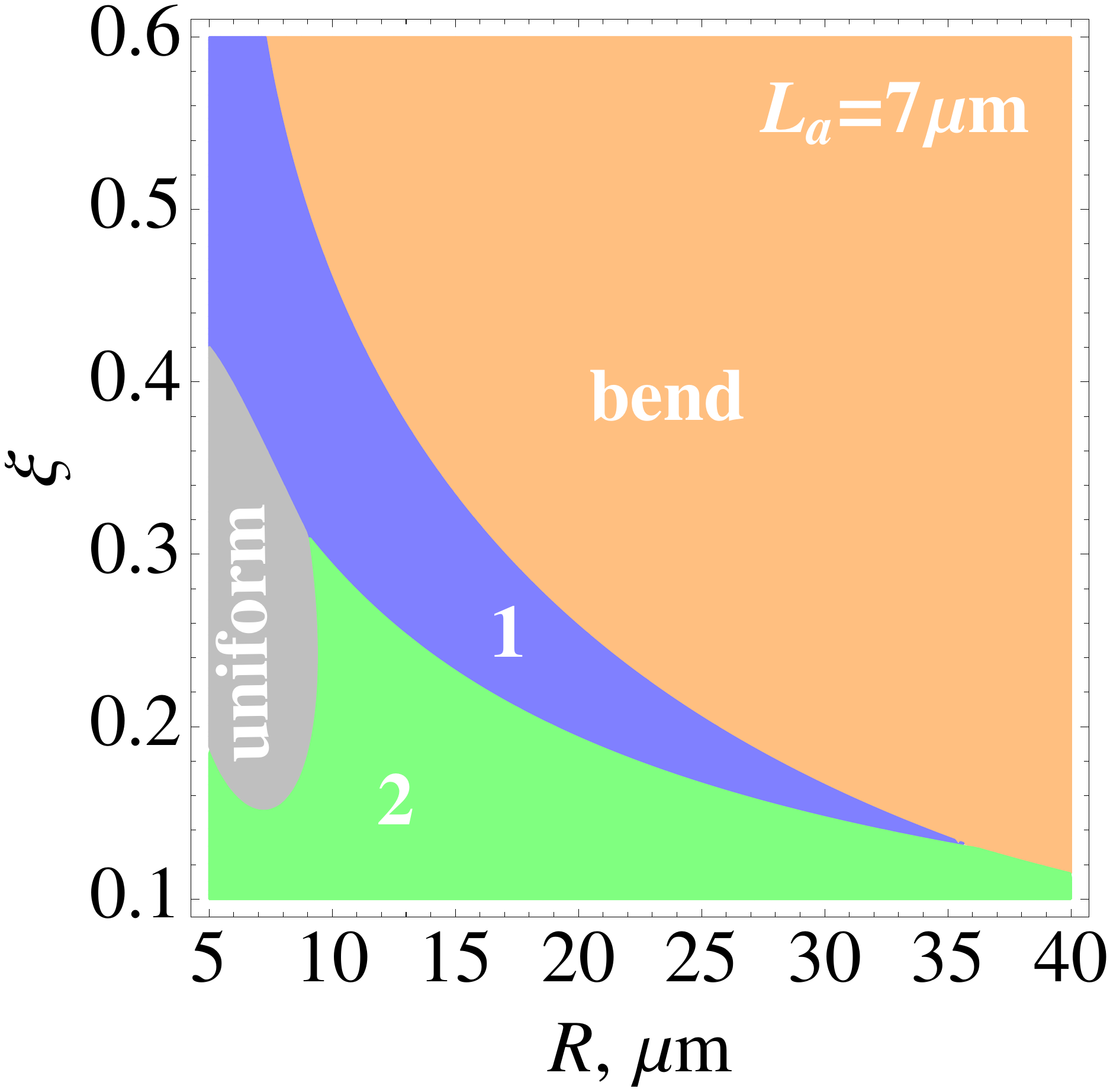}
\caption{\label{fig:aphase} Energetics of the six configurations $\boxed{1}$--$\boxed{6}$ shown in Fig.~\ref{fig:annul}: (a)~The normalized anchoring energy in units of $R/L_a$ as a function of $\xi=R_i/R$. \mcmred{The corresponding expressions are given  in Table~\ref{tab:Ea}; (b)~and (c)~ display the region of parameters where various nematic textures minimize the free energy  (b) is obtained by comparing the {\em elastic} free energy of textures   $\boxed{1}$--$\boxed{3}$ and of the bend state (not that for these configurations the anchoring contribution to the free energy is zero or negligible); (c)~is obtained by comparing the {\em total} free energy of textures $\boxed{1}$--$\boxed{2}$ and of the defect free configuration shown in Fig.~\ref{fig:anchor} for $L_a=7~\mu$m and $\ve=0.5~\mu$m. } }
\end{figure}

We now examine a number of nematic configurations with $k$-fold rotational symmetry, with $k=1,2,3$. These can be obtained by considering different numbers of defect pairs of positive and negative charge sitting at the inner and outer boundaries of the annulus. The total charge is conserved since $\chi=0$~\eqref{eq:conserv}. \mcmred{We consider the textures shown in Fig.~\ref{fig:annul} and evaluate the corresponding energies.   Configurations $\boxed 1$ to $\boxed 3$ can be obtained by starting from the bend state and introducing negative charges $m=-\frac 14$ at the inner boundary and  positive charges $m=+\frac 14$ at the outer boundary of the annulus  and cost less elastic energy from director distortion than $\boxed 7$ to $\boxed 9$, but have a high cost in anchoring energy compared to $\boxed 4$ to $\boxed 6$. Note that textures $\boxed 1$ to $\boxed 3$ and $\boxed 7$ to $\boxed 9$  have the same anchoring energy (see Table~\ref{tab:Ea}). Conversely, the configurations with  $|m|=\frac12$ boundary defects ($\boxed{4}$--$\boxed{6}$) cost more elastic energy associated with curvature of the director field than $\boxed{1}$--$\boxed{3}$, but have lower  anchoring energy, as shown in Fig.~\ref{fig:aphase}a. Charge conservation as given in Eq.~\eqref{eq:conserv} requires that for all  textures considered the anchoring energy density is the same at the inner and the outer boundaries of the annulus. For the configurations  $\boxed4$--$\boxed6$ the anchoring energy is simply $W_aR(1+\xi)\pi\xi^{2n}/2$, with $n=1,2,3$, respectively, for $\boxed4$--$\boxed6$ representing the number of boundary defect pairs. } In the following we exclude the configurations $\boxed{4}$--$\boxed{9}$ from our  analysis, because their elastic/total energy is much larger than that of $\boxed{1}$--$\boxed{3}$, as well as of the defect free states.

In Fig.~\ref{fig:aphase} we compare the energies of the various nematic textures in the annulus. In textures $\boxed{1}$--$\boxed 3$ the $m=\pm\frac 14$ boundary defects have high anchoring energy in fat annuli (small $\xi$) but lower elastic energy than the bend state over a wide range of parameters $\xi$ and $R/\ve$, as shown in  Fig.~\ref{fig:aphase}b which displays the region of stability of the various textures. Thus, for small system size, or weak anchoring, where the elastic energy dominates, we expect the equilibrium textures with one-fold symmetry ($\boxed 1$) to be more favorable in thin annuli and textures with two-fold symmetry ($\boxed 2$) to be favored in thick annuli. The configuration $\boxed 3$ may also be energetically accessible in a narrow range of parameters, consistent with the fact that it is rarely observed in experiments~\cite{thesis:2013}. In all cases boundary-stabilized defect textures are the ground states of confined nematics when $R/L_a\sim O(1)$. In Fig.~\ref{fig:aphase}c, assuming weak anchoring $L_a=7~\mu$m and $\ve=0.5~\mu$m, we illustrate with different colors the minimizers of the total free energy ${\cal F}_{el}+{\cal F}_a$. Defect structures with $m=+\frac 1 4$ are energetically favored, and therefore support our previous estimate of the anchoring extrapolation length. Note that ground states (energy minimizers) with crystalline order on the surface of an embedded torus are also characterized by the presence of positive disclinations on the exterior of the torus and negative disclinations in the interior~\cite{giomi:2008}.

\section*{Concluding remarks}

\mcm{We have shown that, in contrast to what speculated in earlier works~\cite{aarts:2014,thesis:2013,velasco:2014}, a number of defective textures observed experimentally in  nematic films confined to circular and annular geometries can be accounted for within  continuum liquid crystal theory.  By examining the energetics of textures with localized defects at the boundary, and exploiting conservation of topological charge, our work captures the main features of experiments in $fd$-virus~\cite{thesis:2013}  and provides an estimate for the anchoring extrapolation length in these systems --  $L_a\simeq 5~\mu$m. }

Several challenging questions remain unanswered.  Our results suggest that the selection of the symmetry of observed nematic textures is controlled a single dimensionless parameter, $W_a R/K$, and the conservation of topological charge rather than by details of microscopic interactions. But what is the range of validity of the continuum theory, i.e., is it valid when the size of  confinement approaches a few molecular sizes? Another important question concerns the use of the one-elastic constant approximation, which is reasonable for semiflexible polymers, including the $fd$-virus  with its persistence length of the order of the polymer length~\cite{aarts:2012,thesis:2013}, but is not accurate for rigid rods~\cite{delasHeras:2009}.  A significant elastic anisotropy was measured experimentally for systems composed of tobacco mosaic virus~\cite{fraden:1985} with bend constant larger than splay constant, $K_3\simeq 17 K_1$. On the contrary, according to theoretical predictions~\cite{williams:1994} the splay elastic constant $K_1$ diverges as the length of the molecule while the bend elastic constant $K_3$ depends on the rigidity of the molecules. Thus for (semi)flexible polymers one expects $K_1> K_3$. It would therefore be of interest  to quantify the energetics of topological defects in case of strong anisotropy of the elastic constants. The authors~\cite{rey:2002} analyzed the influence of splay-bend anisotropy on the formation of fiber texture in discotic liquid crystals with fixed boundary conditions and topological defects in the bulk. Our approach could also be generalized to other planar geometries with non-monotonic curvature of the boundary, such as the square plates examined in~\cite{thesis:2013,aarts:2014}.  \mcm{Finally it is well known that in three dimensions topological defects affect the shape of nematic droplets~\cite{schoot:2003,lavrent:2013}. It would similarly be interesting to examine the interplay between defect textures and shape in thin nematic films confined by a deformable boundary.}

\subsection*{Acknowledgements}
 We thank  Jos\'e Alvarado for valuable discussions. This work was  supported by the Syracuse University Soft Matter Program. MCM acknowledges support from the National Science Foundation (NSF) through award DMR-1305184. KBL was supported by an NSF-IGERT traineeship through award DGE-1068780. MJB and OVM thank ICERM at Brown University for hospitality during the completion of this work.%%  and acknowledge financial support through the grant NSF-DMS-0931908. 

\footnotesize{
\bibliographystyle{kp}
\bibliography{ref_arxiv}
}

\end{document}